\documentclass[preprint,12pt]{elsarticle}
\pdfoutput=1

\usepackage{amsmath}
\usepackage{amstext}
\usepackage{amssymb}
\usepackage{lineno}
\usepackage{makecell}
\usepackage{color}
\usepackage{soul}
\usepackage[utf8]{inputenc}
\usepackage{fancyhdr}
\usepackage[normalem]{ulem}

\usepackage[x11names]{xcolor}
\usepackage[colorlinks]{hyperref}

\AtBeginDocument{\hypersetup{citecolor=DodgerBlue4}}

\modulolinenumbers[5]

\biboptions{sort&compress}

\AtBeginDocument{\hypersetup{citecolor=Green3,linkcolor=Red1,urlcolor=DeepPink2}}
\begin{document}
\begin{frontmatter}
\title{Sterile neutrinos with altered dispersion relations revisited}

\author[IFIC,Dep]{G. Barenboim}
\ead{Gabriela.Barenboim@uv.es}
\author[IFIC,Dep]{P. Mart\'inez-Mirav\'e} 
\ead{pamarmi@ific.uv.es}
\author[IFIC]{C. A. Ternes} 
\ead{chternes@ific.uv.es}
\author[IFIC,Dep]{M. T\'ortola} 
\ead{mariam@ific.uv.es}

\address[IFIC]{Instituto de F\'{i}sica Corpuscular, CSIC-Universitat de Val\`{e}ncia, 46980 Paterna, Spain}
\address[Dep]{Departament de Física Teòrica, Universitat de València, 46100 Burjassot,  Spain}

\vspace{10pt}

%\begin{indented}
%\item[] bla
%\end{indented}

\begin{abstract}
In this paper we investigate neutrino oscillations with altered dispersion relations in the presence of sterile neutrinos. Modified dispersion relations represent an agnostic way to parameterize new physics. Models of this type have been suggested to explain global neutrino oscillation data, including deviations from the standard three-neutrino paradigm as observed by a few experiments. We show that, unfortunately, in this type of models new tensions arise turning them incompatible with global data.
\end{abstract}

\begin{keyword}
Neutrino oscillations\sep BSM physics\sep Lorentz violation 
\end{keyword}
\end{frontmatter}

\fancyhf{}
\renewcommand{\headrulewidth}{0pt}
\thispagestyle{fancy}

\section{Introduction}
\label{sec:intro}
Over the last approximately 20  years, neutrino oscillation measurements have become 
more and more precise and are now entering the precision era. Most of the current 
data coming from experiments using neutrinos from the Sun, reactors, the atmosphere 
and particle accelerators can be described in terms of three-neutrino oscillations, 
which depend on  six oscillation parameters: two mass splittings ($\Delta m_{31}^2$, 
$\Delta m_{21}^2$), three mixing angles ($\theta_{12}$, $\theta_{13}$ and $\theta_{23}$)
and a CP-violating phase ($\delta$). Many of these parameters are measured very well as 
of now~\cite{deSalas:2017kay}. The remaining unknowns in this picture are the exact value of 
the CP-phase $\delta$, the octant of the atmospheric angle ($\sin^2\theta_{23}<0.5$ or 
$\sin^2\theta_{23}>0.5$) and the neutrino mass ordering ($\Delta m_{31}^2>0$ or 
$\Delta m_{31}^2<0$). 
The most recent oscillation data already provide some hints in favor of maximal CP violation and second octant of $\theta_{23}$, as well as a clear preference (above the  3$\sigma$ level) for the normal mass ordered neutrino spectrum\cite{deSalas:2017kay}, although they are not fully conclusive yet.
Note that,  combining oscillation data with recent cosmological observation results, a 3.5$\sigma$ preference for normal ordering can be obtained~\cite{Gariazzo:2018pei,deSalas:2018bym}. 

Beyond the standard three-neutrino scenario, currently well established and characterized, some observations 
might suggest the existence of a fourth neutrino mass eigenstate. In the 90s, the LSND experiment observed the
appearance of electron antineutrinos in a muon antineutrino 
beam~\cite{Athanassopoulos:1996wc,Athanassopoulos:1997pv,Aguilar:2001ty}. 
A similar signal was recently observed in the MiniBooNE experiment~\cite{Aguilar-Arevalo:2018gpe}. 
Anomalies have also been observed  in the electron (anti-) neutrino disappearance channel, known as the Gallium 
anomaly~\cite{Abdurashitov:2009tn,Laveder:2007zz,Acero:2007su,Giunti:2010zu} and 
the reactor antineutrino anomaly~\cite{Mention:2011rk}. 
The common feature of all these anomalous results is their short baseline, or $L/E$ of order 1 km/GeV and, therefore, all of them can be explained in terms of a fourth sterile neutrino with $\Delta m_{41}^2\approx 1$ eV$^2$, see 
for example Ref.\cite{Gariazzo:2015rra}.
However, with new data coming from different long baseline experiments~\cite{Adamson:2017uda,TheIceCube:2016oqi,Aartsen:2017bap,Albert:2018mnz,Abe:2019fyx} a tension between the results observed in muon neutrino beams at  disappearance and  appearance channel arises, 
see Refs.~\cite{Gariazzo:2017fdh,Dentler:2018sju}.
The reason is that the mixing angles and mass splittings required to explain the short baseline anomalies should produce a visible 
effect at the long baseline sector, that is absent. Therefore, the simplest 3+1 scheme can not 
explain all the data simultaneously . For current reviews on this topic see Refs.~\cite{Giunti:2019aiy,Boser:2019rta}. 
It has been shown that adding simply more sterile neutrinos will not resolve this tension either~\cite{Giunti:2015mwa}.  

This hot topic has been addressed in many articles since the latest results from MiniBooNE 
appeared~\cite{Bertuzzo:2018itn,Bertuzzo:2018ftf,Ballett:2018ynz,Doring:2018cob,Liao:2018mbg,Denton:2018dqq,Jordan:2018qiy,Arguelles:2018mtc,deGouvea:2019qre,Dentler:2019dhz,Giunti:2019sag}.
As one can see, many theories are being tested, some of which are directly related  to neutrino
oscillations as in Refs.~\cite{Liao:2018mbg,Denton:2018dqq,deGouvea:2019qre,Dentler:2019dhz,Giunti:2019sag}.

In this paper, we focus on neutrino oscillations with altered dispersion relations (ADR)~\cite{Pas:2005rb,Hollenberg:2009ws}. Modified dispersion relations are an economic and agnostic way to encompass a whole bunch of new physics models. 
Using the fact that neutrino oscillation experiments are nothing but a fancy interferometer (neutrinos are produced as flavour eigenstates but propagate as mass eigenstates), we can use them to study effects that 
would be otherwise too small to be observed, like Lorentz violation~\footnote{Let us remind the reader that, even if modified dispersion relations imply that Lorentz symmetry is broken, the theory is invariant under changes of coordinates.}.
As it is well known in the Standard Model, the (scalar) Higgs field acquires a vacuum expectation value (\textit{vev}) breaking the electroweak symmetry and giving masses to fermions.  Therefore, it won't be surprising that in
string theory (or in quantum gravity) not a scalar but a tensor field would be the one acquiring a \textit{vev}. As a result, the interaction of the fields that couple to these \textit{vev}, which can be thought of as background fields,  will be velocity and direction dependent. In other words, these \textit{vev} will trigger the breakdown of Lorentz symmetry.

Of course, Lorentz violations  can arise naturally  also in theories with extra 
dimensions~\cite{ArkaniHamed:1998rs,Antoniadis:1998ig,ArkaniHamed:1998nn,ArkaniHamed:1998vp}. 
In this type of theories~\cite{Pas:2005rb,Marfatia:2011bw,Doring:2018ncz}, sterile neutrinos can travel through the extra dimensions, causing a resonant oscillation behavior for a certain energy range, which might give an explanation for the anomalies observed in a few experiments~\cite{Hollenberg:2009qf,Hollenberg:2009ak}, without getting into conflict with cosmological observations~\cite{Aeikens:2016rep}, which is not the case for a scenario with simple sterile neutrinos~\cite{Gariazzo:2019gyi}.
The resonant behavior is a key ingredient in the set-up, as it allows to tune the energy range where the effect triggers and
guarantees that it is set-off outside of this range.
It has been argued~\cite{Doring:2018cob}, that these models do not affect the results obtained by the long baseline experiments. However, here we show that the parameters needed to produce sizeable effects in short baseline oscillations, indeed do spoil the oscillation probabilities in other neutrino oscillation experiments and, therefore, do not give a solution to the tension observed in short baseline oscillations. 

 Our paper is structured as follows: In Sec.~\ref{sec:3+1} we first give a brief introduction to 3+1 mixing. Then, we discuss ADRs in this scenario, where we consider intrinsic ADRs and ADRs coming from an effective potential affecting neutrino propagation. In Sec.~\ref{sec:3+3} we extend this discussion to the case of three sterile neutrinos and address the consistency of its predictions. Finally, in Sec.~\ref{sec:conc} we draw our conclusions.

%%%%%%%%%%%%%%%%%%%%%%%%%%%%%%%%%%%%%%%%%%%%%%%%%%%%%%%%%%%%%%%%%%%%%%%%%%%%%%%%%%%%%%%%%%%%%%%%%%%%%%%%%%%%%%%%%%%%%%%%%%%%%%%%%%%%%%%%%%%%%%%%%%%%%%%%%%%%%%%%%%%%%%%%%%%%%%%%%%%%%%%%%%%%%

\section{Altered dispersion relations in a 3+1 scenario}
\label{sec:3+1}

In order to explain the anomalies mentioned in the introduction, the existence of a fourth neutrino was suggested. This additional neutrino must be sterile, hence a Standard Model gauge singlet, or heavy enough to avoid bounds by LEP on the number of active neutrino families~\cite{ALEPH:2005ab}. In this paper, we will consider only light sterile neutrinos. In this case, the lepton mixing matrix has to be extended, adding three new angles, two new phases and a new mass splitting. In the simplest scenario, the Hamiltonian describing the neutrino propagation in matter is given by
\begin{equation}
H = \frac{1}{2E} U 
\begin{pmatrix}
    m_1^2 & 0 & 0 & 0  \\
    0 & m_2^2 & 0 & 0  \\
    0 & 0 & m_3^2 & 0  \\
    0 & 0 & 0 &  m_4^2 
\end{pmatrix}
    U ^{\dagger} 
+ \begin{pmatrix}
    V_{\text{CC}} & 0 & 0 & 0  \\
    0 & 0 & 0 & 0  \\
    0 & 0 & 0 & 0  \\
    0 & 0 & 0 &  -V_{\text{NC}}
\end{pmatrix}\,.
\end{equation}
Here, $E$ is the neutrino energy, $U$ 
is the matrix describing neutrino mixing, $m_i$ 
are the neutrino masses and $V_\text{CC}$ and 
$V_\text{NC}$ are the charged current (CC) and
neutral current (NC) potentials, respectively. Note that, since sterile neutrinos do not feel the weak interaction, the neutral current component of the potential, $V_\text{NC}$, can not be eliminated from the expression of the effective potential, as it happens in the standard three-neutrino case.
The neutrino mixing  is now parameterized in terms of the $4\times4$ unitary matrix 
\begin{equation}
U = \tilde{U}_{34} U_{24} \tilde{U}_{14} U_{23} \tilde{U}_{13} U_{12}\,,
\end{equation}
where the matrix $U_{ij}$ represents a
rotation in the $i$-$j$ plane and the tilde indicates that the corresponding mixing angle is accompanied by a CP-phase. Since in this work we are interested in effects occurring only on short baselines or in channels which are not sensitive to matter effects, we will not consider the matter potential here. As argued in the introduction, this simplest extension can not explain the anomalous results obtained by a few experiments without being in tension with other experiments. In this section we consider two possible extensions of this model. 

\subsection{Intrinsic modified dispersion relation}
The excess of events found in MiniBooNE can be studied assuming the existence of a sterile neutrino with $\Delta m^2_{41} \simeq 1~ \text{eV}^2$. In that case, the electron neutrino appearance probability is given by
\begin{equation}
 P_{\mu e} \simeq \sin^2 \theta_{24} \sin^2 2\theta_{14} \left( \frac{\Delta m ^2 _{41} L}{4E} \right) .
\end{equation}

The existing bounds on $\theta_{24}$ come mainly from the non-observation of a signal of sterile neutrinos in the disappearance channel
in MINOS/MINOS+~\cite{MINOS:2016viw,Adamson:2017uda} and IceCube~\cite{TheIceCube:2016oqi}. At first approximation, the disappearance $\nu_\mu$ oscillation probability in the 3+1 scheme is given by
\begin{equation}
 P_{\mu\mu} \simeq 1 - \sin^2 2 \theta_{23} \cos 2 \theta_{24}\sin^2 \left( \frac{\Delta m ^2 _{31} L}{4E} \right) - \sin^2 2\theta_{24} \sin^2 \left( \frac{\Delta m ^2 _{41} L}{4E} \right) .
 \label{eqn:mu-mu}
\end{equation}
In the case of MINOS/MINOS+ and IceCube, since the kinematic phase $\frac{\Delta m ^2 _{41} L}{4E}$ is very large, the last term 
in the expression above is considered to be averaged to $1/2 \sin ^2 2 
\theta_{24}$. The strong bounds on the 3+1 scenario coming from these experiments undermine the explanation of the anomalies in terms of a sterile neutrino. However, it has been claimed that altered dispersion relations could relax the tension between appearance and disappearance experiments.

\begin{figure}[ht!]
    \centering
    \includegraphics[width = 0.9\textwidth]{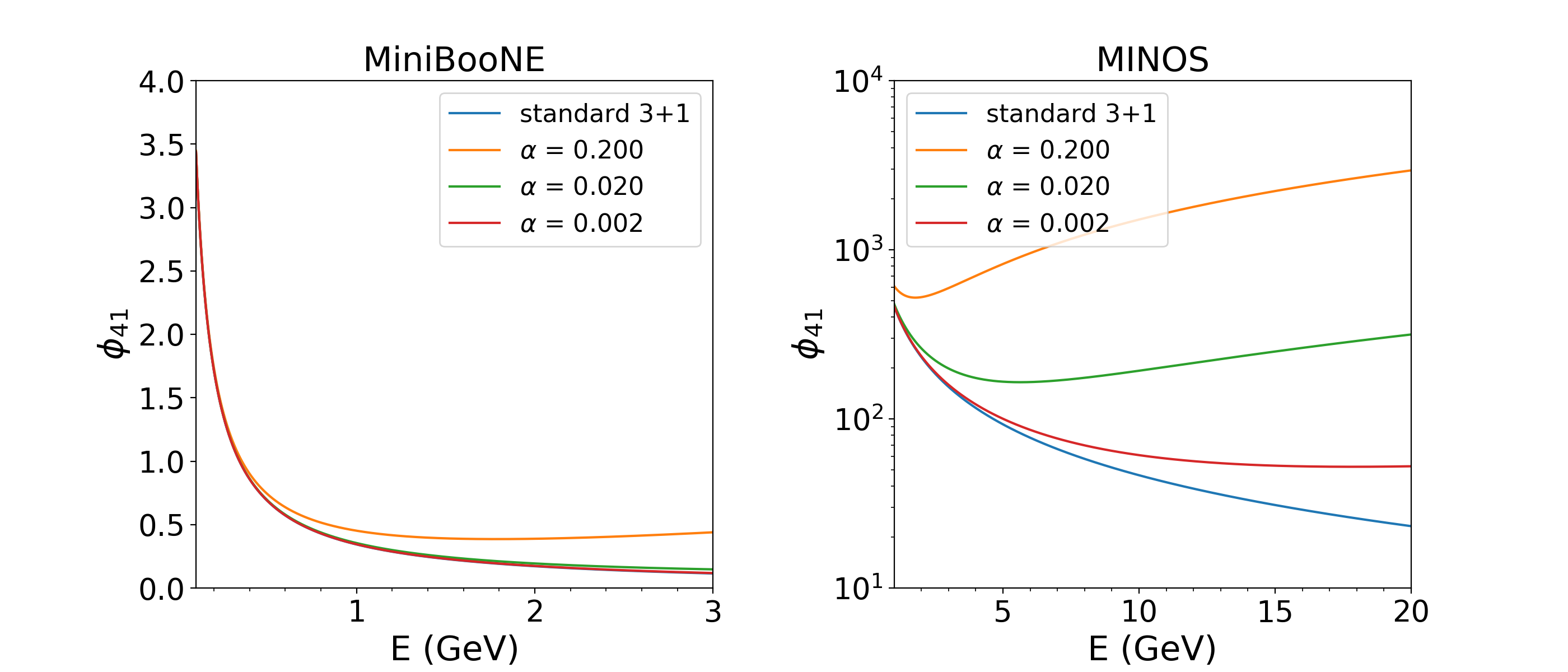}
    \caption{The kinematic phase $\phi_{41}$ as a  function of the energy for MiniBooNE ($L = 0.541$ km) and MINOS ($L = 731$ km) for $\Delta m^2_{41} = 1.4~\text{eV}^2$. Different values of $\alpha$ are also presented. }
    \label{fig:beta1}
\end{figure}

A modification of the dispersion relation occurs when the energy momentum relation $E^2 = p^2 + m^2$ does not hold any more. Alterations of this type can appear in theories with Lorentz violation~\cite{Kostelecky:2003cr, Diaz:2011ia, Kostelecky:2011gq, Barenboim:2018ctx}. Here we will assume a generic Lorentz violating term associated to the fourth mass eigenstate. In this case, the kinematic phase changes  according to 
\begin{equation}
      \phi_{4i} = \frac{\Delta m_{4i}^2 L}{4E}   \longrightarrow  
\phi_{4i}=\left(\frac{\Delta m_{4i} ^2}{4E} + f(E)\right)L  \, , \quad \mathrm{with} \quad i = 1,2,3 \,.
\end{equation}
For simplicity, we choose 
$f(E) = \alpha E$. If the function $f(E)$ 
is positive ($\alpha > 0$), the kinematic
phase is larger than its corresponding value in 
the 3+1 neutrino standard framework, as it is 
shown in Figure \ref{fig:beta1}. This translates 
in the fact that probability terms controlled 
by $\Delta m^2 _{41}$ get smeared out at smaller
energies. Such a behavior has no impact on the
bounds set by MINOS/MINOS+ and IceCube on 
$\theta_{24}$, since the term depending on 
$\phi_{41}$ is already averaged to 1/2. Adding 
a modified dispersion relation that makes 
the kinematic phase grow with the energy would 
only result on this term getting averaged to 1/2 
at a lower energy.
\begin{figure}[ht!]
    \centering
    \includegraphics[width = 0.9\textwidth]{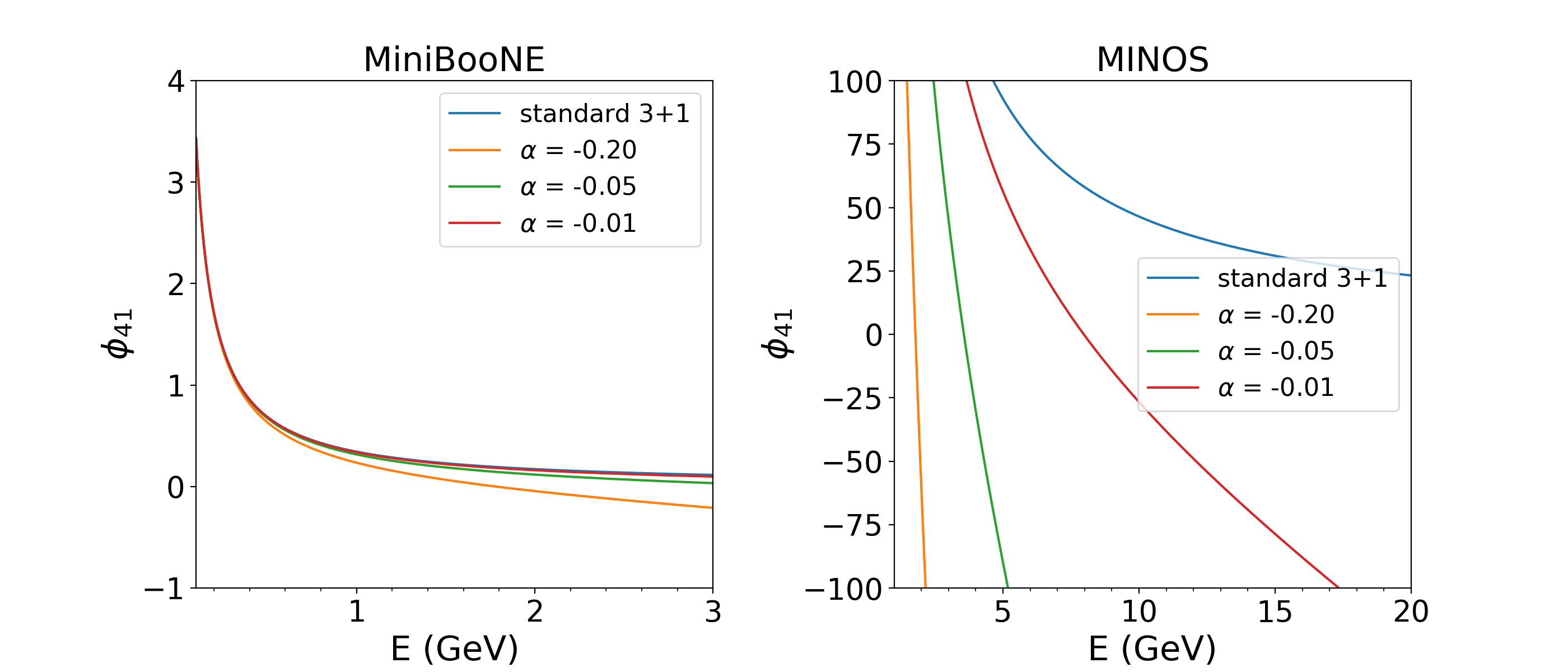}
    \caption{The kinematic phase $\phi_{41}$ as a function of the energy for MiniBooNE ($L = 0.541$ km) and MINOS ($L = 731$ km). Different values of $\alpha$  are also presented. }
    \label{fig:multiplebeta}
\end{figure}
If the function $f(E)$ were negative 
($\alpha < 0$), the kinematic phase could eventually reach very small values and even get to zero. In that case,
\begin{equation}
 P_{\nu_\mu \rightarrow \nu_\mu} \simeq 1 - \sin^2 2 \theta_{23} \cos 2\theta_{24}\sin^2 \Big( \frac{\Delta m ^2 _{31} L}{4E} \Big)  - \sin^2 2\theta_{24} \phi_{41}^2 .
\end{equation}
Then, we can conclude that  a very small kinematic phase $\phi_{41}$ along the energy range of MINOS could weaken the bounds on $\theta_{24}$, since the bound would no longer apply to $\frac{1}{2}\sin^2 2\theta_{24}$ but to $\sin^2 2\theta_{24}\phi_{41}^2$. 
Indeed, one can always choose a modified dispersion relation $f(E)$ such that, for a given energy $E_{0}$ in the spectrum of MINOS, $f(E_0) = \Delta m _{41}^2/4E_0$ and, then, $\phi_{41}(E_0) = 0$. 
Thus, if, along the energy spectrum of MINOS, $\phi_{41}$ were very small, it would be possible to weaken its bounds on the 3+1 framework, as it was previously explained. However, the condition of $\phi_{41}$ being small would be valid only for a small interval of the energy spectrum. After a certain value of the energy, the modulus of the kinematic phase $\phi_{41}$ becomes very large and the 3+1 neutrino picture is recovered, with a contribution from the sterile neutrino  to the appearance probability equal to $1/2 \sin^2 2 \theta_{24}$.
This behavior is illustrated in Figure \ref{fig:multiplebeta}.
 Different parametrisations of the modified dispersion relation $f(E)$ with  physically reasonable energy dependencies have been  explored. However, no substantial difference in the argumentation merits to be reported.
Therefore, modified dispersion relations whose origin is the violation of Lorentz invariance, together with sterile neutrinos, can not reconcile the LSND and MiniBooNE anomalies with the results of other atmospheric and long baseline experiments, since the tension reported in Refs.~\cite{Gariazzo:2017fdh,Dentler:2018sju} is not alleviated.
 Note that, apart from the problems mentioned here, neutrinos in this scenario would be superluminal, giving rise to further problems not discussed in this paper.

%%%%%%%%%%%%%%%%%%%%%%%%%%%%%%%%%%%%%%%%%%%%%%%%%%%%%%%%%%%%%%%%%%%%%%%%%%%%%%%%%%%%%%%%%%%%%%%%%%%%%%%%%%%%%%%%%%%%%%%%%%%%%%%%%%%%%%%%%%%%%%%%%%%%%%%%%%%%%%%%%%%%%%%%%%%%%%%%%%%%%%%%%%%%%%%%%%%%%%%%%%%%%%%

\subsection{Modified dispersion relations from effective potentials}

Altered dispersion relations due to effective potentials in the Hamiltonian can lead to energy dependent oscillation parameters. The nature of such an additional potential can be shortcuts through extra dimensions~\cite{Pas:2005rb,Doring:2018ncz}. This type of modification, together with the existence of one or more sterile neutrinos, has been suggested as a solution to the anomalies found in neutrino oscillation experiments~\cite{Doring:2018ncz,Doring:2018cob}. 
In the 3+1 scenario with modified dispersion relations, the neutrino Hamiltonian in vacuum is given by
\begin{equation}
H = \frac{1}{2E} U 
\begin{pmatrix}
    m_1^2 & 0 & 0 & 0  \\
    0 & m_2^2 & 0 & 0  \\
    0 & 0 & m_3^2 & 0  \\
    0 & 0 & 0 &  m_4^2 
\end{pmatrix}
    U ^{\dagger} 
- \begin{pmatrix}
    0 & 0 & 0 & 0  \\
    0 & 0 & 0 & 0  \\
    0 & 0 & 0 & 0  \\
    0 & 0 & 0 &  \epsilon E ^{\alpha} 
\end{pmatrix}\,.
\end{equation}
This type of effective potential was initially proposed for $\alpha = 1$, which would correspond to sterile neutrinos traveling through extra dimensions \cite{Pas:2005rb, Doring:2018cob}. The parameter $\epsilon$ is related to the time difference between the active and the sterile neutrino traveling through extra dimensions. 

\begin{figure}[t!]
    \centering
    \includegraphics[width = 0.8\textwidth]{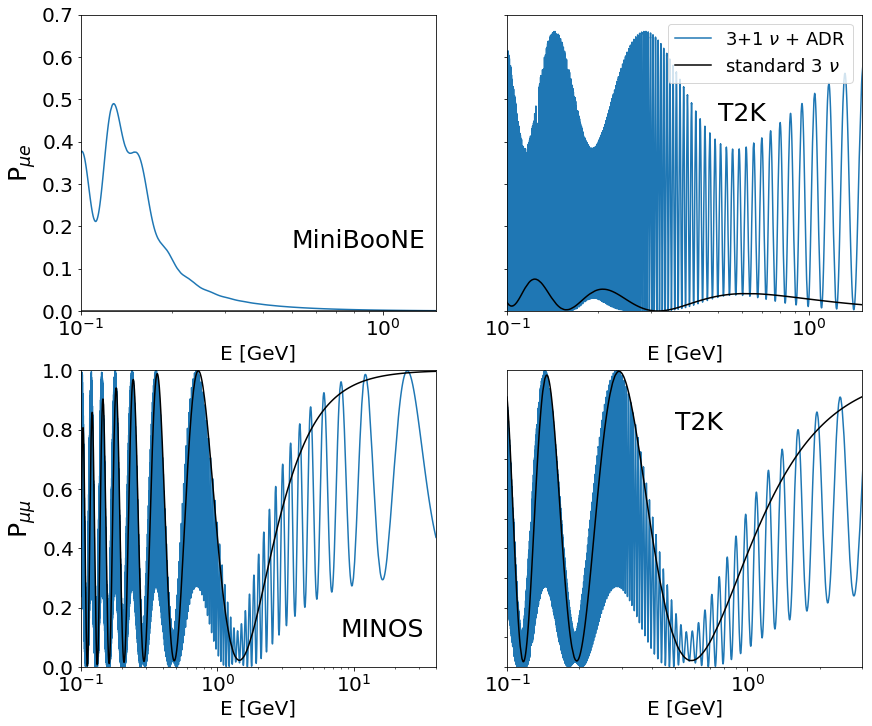}
    \caption{Comparison between the predictions of the three neutrino standard picture and a 3+1 model with altered dispersion relations with $\epsilon = 5\times 10^{-17}$. The upper panels show the appearance probability at MiniBooNE ($L=0.541$ km) (left) and T2K ($L=295$ km) (right), while the lower panels show the disappearance probability at MINOS ($L=731$ km) (left) and T2K ($L=295$ km) (right).}
    \label{fig:3+1}
\end{figure}

It is clear that such a potential induces energy dependencies in the oscillation parameters. The value of the parameter $\alpha$, which is model dependent, sets how wide or narrow the resonant effect is. It is important to notice that both mixing angles and mass splittings are now energy dependent. The latter ones can be obtained from the eigenvalues of the Hamiltonian, $\lambda_i$, as $m_{i, \text{eff}}^2 = 2  E \lambda_i (E)$.
In principle, the resonant behavior of this scenario could relax the tension in data coming from appearance (MiniBooNE, essentially) and disappearance experiments (principally MINOS/MINOS+). 
In order to be consistent with Ref.~\cite{Doring:2018cob}, we adopt the untypical parameterization 
\begin{equation}
 U = U_{23} U_{13} U_{12} U_{14}\,. 
\end{equation}
In this notation only one mixing angle is needed to
induce a non-zero short baseline appearance 
channel and not two mixing angles as in the 
standard case.
The electron appearance probability in MiniBooNE is given by
\begin{equation}
     P_{\mu e} \simeq 4 |U_{e4}^\text{eff}|^2 |U_{\mu4}^\text{eff}|^2\sin^2\left(\frac{\Delta m ^{2,\text{eff}}_{41} L}{4E}\right),
\end{equation}
where $U^{\text{eff}}$ and $\Delta m^2_{\text{eff}}$ are the corresponding effective mixing matrix and mass splitting once the additional effective potential is considered.
Therefore, if the combination 
$ 4 |U_{e4}^\text{eff}|^2 |U_{\mu4}^\text{eff}|^2$ 
happens to be large at the energy of the MiniBooNE 
anomalous signal ($E \leq  0.3$ GeV), this 
mechanism could give rise to a significant 
appearance probability in MiniBooNE. 
Unfortunately, forcing 
$4 |U_{e4}^\text{eff}|^2 |U_{\mu4}^\text{eff}|^2$ 
to be large  also affects the oscillation 
probabilities at long baseline experiments. In the 
upper left panel of Fig.~\ref{fig:3+1} we plot the 
oscillation probability for MiniBooNE showing the 
required resonance at the energies of interest, as 
indicated by the blue curve. However, in the case 
of MINOS/MINOS+ (lower left panel) we see new fast 
oscillations, which should in average lower the 
signal rate in the disappearance channel with 
respect to the standard case (black line). The same 
happens in the disappearance channel at the T2K 
experiment. Most striking, however, is the expected 
signal at the appearance channel of T2K. There, one 
can see a very fast oscillation pattern reaching 
very large oscillation probabilities. This is due 
to the fact that the neutrino energy ranges covered 
by MiniBooNE and T2K overlap and, therefore, an 
energy dependent excess in MiniBooNE should  have a 
visible effect in T2K as well. 
The standard oscillation parameters used to create 
these plots are those from Tab.~\ref{tab:oscparam}.
For the new parameters we choose 
$\sin^2\theta_{14} = 0.05 $, 
$\Delta m_{41}^2 = 1.59$~eV$^2$ and $\epsilon = 5 \times 10^{-17}$~\footnote{We have checked different combinations of oscillation parameters, obtaining always  the same qualitative result.}.
Note that using different values for $\delta$ would 
have no effect in MiniBooNE and leave also the 
T2K disappearance probability unchanged , 
while producing only a slight 
modification in the T2K  appearance probability.

\begin{table}[!t]
\centering
  \catcode`?=\active \def?{\hphantom{0}}
   \begin{tabular}{|c|c|}
    \hline
    Parameter & Value
    \\
    \hline
    $\Delta m^2_{21}$       & $7.55\times 10^{-5}$ eV$^2$\\  
    $\Delta m^2_{31}$       &  $2.50\times 10^{-3}$ eV$^2$\\
    $\sin^2\theta_{12}$     & 0.32\\ 
    $\sin^2\theta_{23}$     & 0.547\\
    $\sin^2\theta_{13}$     & 0.0216\\
    $\delta$                & 0\\
    \hline
    \end{tabular}
    \caption{The standard neutrino oscillation parameters used in the
      analysis, taken from Ref.~\cite{deSalas:2017kay}, except for $\delta$ which is set to zero for simplicity.}
    \label{tab:oscparam} 
\end{table}

Previous studies~\cite{Doring:2018cob} have pointed out an additional source of inconsistencies with the experimental data. In particular, it has been shown that, for energies above the resonance, and as a consequence of the energy dependence of the effective mass eigenstates,  atmospheric neutrino experiments  should also have presented clear deviations from the three-neutrino picture. Nonetheless, in a 3+3 scheme this can be (unfortunately only) partially solved.

%%%%%%%%%%%%%%%%%%%%%%%%%%%%%%%%%%%%%%%%%%%%%%%%%%%%%%%%%%%%%%%%%%%%%%%%%%%%%%%%%%%%%%%%%%%%%%%%%%%%%%%%%%%%%%%%%%%%%%%%%%%%%%%%%%%%%%%%%%%%%%%%%%%%%%%%%%%%%%%%%%%%%%%%%%%%%%%%%%%%%%%%%%%%%%%%%%%%%%%%%%%%%%%

\section{Altered dispersion relations in a 3+3 scenario}
\label{sec:3+3}

Given the impossibility to reconcile the neutrino anomalous results in the context of a 3+1 scenario with altered dispersion relations,  some works have proposed an alternative explanation in terms of a 3+3 scenario with or without extra new physics.
If we consider three sterile neutrinos, our mixing matrix has to be further extended. The full mixing matrix is now given by~\cite{Xing:2011ur}
\begin{equation}
U = U_{36} U_{26} U_{16} U_{35} U_{25} U_{15} U_{34} U_{24} U_{14} U_{23} U_{13} U_{12}\,,
\end{equation}
where we ignored possible CP-phases. The alteration of the dispersion relations can be introduced through an effective neutrino potential given by~\cite{Doring:2018cob}
\begin{equation}
 V_{\text{eff}} = -
\begin{pmatrix}
0 & 0 & 0 & 0 & 0 & 0 \\
0 & 0 & 0 & 0 & 0 & 0 \\
0 & 0 & 0 & 0 & 0 & 0 \\
0 & 0 & 0 &  \epsilon E & 0 & 0 \\
0 & 0 & 0 & 0 & \kappa E& 0 \\
0 & 0 & 0 & 0 & 0 & \xi E  
\end{pmatrix}\, ,
\label{eq:V3+3}
\end{equation}
where we introduce three new parameters $\epsilon, \kappa$ and $\xi$. This potential can be easily generalized by changing the power of the energy dependence. 
A resonant-like effect induced by this potential in MiniBooNE would require positive values for the coefficients $\epsilon$, $\kappa$ and $\xi$. 

Note that the initial proposal of the model in Ref.~\cite{Doring:2018cob} uses an unconventional parametrization of the mixing matrix, 
\begin{equation}
U = U_{23} U_{13} U_{12} \hat{U}_{14} \hat{U}_{25} \hat{U}_{36}\,, 
\end{equation}
with only three new mixing angles: $\hat{\theta}_{14}$, $\hat{\theta}_{25}$, $\hat{\theta}_{36}$. Moreover, they are imposed to be equal, $\hat{\theta}_{14} = \hat{\theta}_{25} = \hat{\theta}_{36} = \theta$. The reason for this shall be explained below. Since the original parametrization~\cite{Doring:2018cob} is easier to handle for the discussion we are going to present, we will use it from now on. For our numerical studies we will use again the standard oscillation parameters from Tab.\ref{tab:oscparam} and the new parameters from Tab.~\ref{tab:newparam}, for which we use two different sets~\footnote{Note that we did not restrict our analysis only to these two sets, but tried to cover all the possibilities leading to a significant signal in MiniBooNE and LSND. We found the general trend to be similar to the one presented here.}. For simplicity, we have set all of the CP-phases to zero. 
\begin{table}[!h]
\centering
  \catcode`?=\active \def?{\hphantom{0}}
   \begin{tabular}{|c|c|c|}
    \hline
    parameter & set 1 & set 2 
    \\
    \hline
    $\Delta m^2_{41}$   &   $1.59 $ eV$^2$         &     $1.59 $ eV$^2$\\
    $\sin^2\theta$      &   $0.05$                 &     $0.05$\\
    $\epsilon$          &   $8 \times 10^{-16}$    &     $5 \times 10^{-15}$\\
    $\kappa$            &   $4 \times 10 ^{-17}$  &     $5 \times 10^{-17} $\\
    $\xi$               &   $4 \times 10^{-17}$   &     $5 \times 10^{-17}$\\
    \hline
    \end{tabular}
    \caption{New  oscillation parameters used in the analysis, with $\theta=\hat{\theta}_{14} = \hat{\theta}_{25} = \hat{\theta}_{36}$.}
    \label{tab:newparam} 
\end{table}

We choose $\Delta m_{41}^2 = 1.59$~eV$^2$. The other new mass differences are chosen to be
\begin{align}
\Delta m^2 _{51} &= \Delta m^2 _{41} + \Delta m^2 _{21}\,,\nonumber\\
\Delta m^2 _{61} &= \Delta m^2 _{41} + \Delta m^2 _{31}\,.
\end{align}
These choices made in the initial proposal can potentially help to deal with the inconsistencies related to the values of the mass splittings at energies above the resonance. The idea behind it is that, above the resonance, sterile and active neutrinos swap their roles and the active-to-sterile mixing is suppressed. Then, the mass differences $\Delta m^2_{54}$ and $\Delta m^2_{64}$ are the ones accounting for the oscillations measured experimentally, $\Delta m^2_{21}$ and $\Delta m^2_{31}$, respectively. Consequently, at high energies they have to be equal to $\Delta m^2 _{21}$ and $\Delta m^2_{31}$. Choosing $\hat{\theta}_{14} = \hat{\theta}_{25} = \hat{\theta}_{36}$ is necessary in order not to spoil this behavior at high energies. 

\begin{figure}[t!]
    \centering
    \includegraphics[width = 0.8\textwidth]{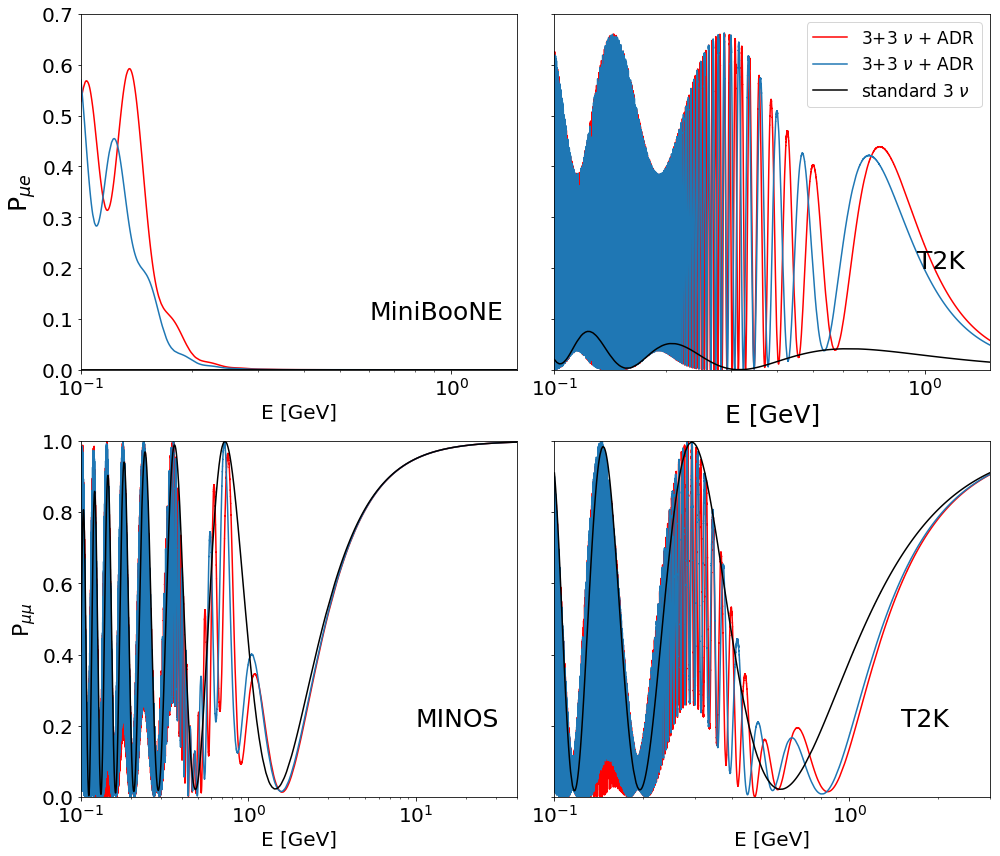}
    \caption{Comparison between the predictions of the three neutrino standard picture and a 3+3 model with altered dispersion relations for the two sets of parameters mentioned in the main text.}
    \label{fig:3+3}
\end{figure}

Unfortunately, the tension between T2K and MiniBooNE arising from the energy dependence of the mixing angles is still present in models with altered dispersion relations and three sterile neutrinos. As in the 3+1 case, it is possible to achieve the desired resonant effect in MiniBooNE, see the upper left panel of Fig.~\ref{fig:3+3}. This time also the MINOS/MINOS+ probability reproduces the standard one much better, since the fast oscillations appear only for rather low energies. However, it is clear that, as in the case of the 3+1 scenario, the oscillation probabilities at T2K are spoiled, as shown in the right panels of Fig.~\ref{fig:3+3}.

\begin{figure}[t!]
    \centering
    \includegraphics[width= 0.9\textwidth]{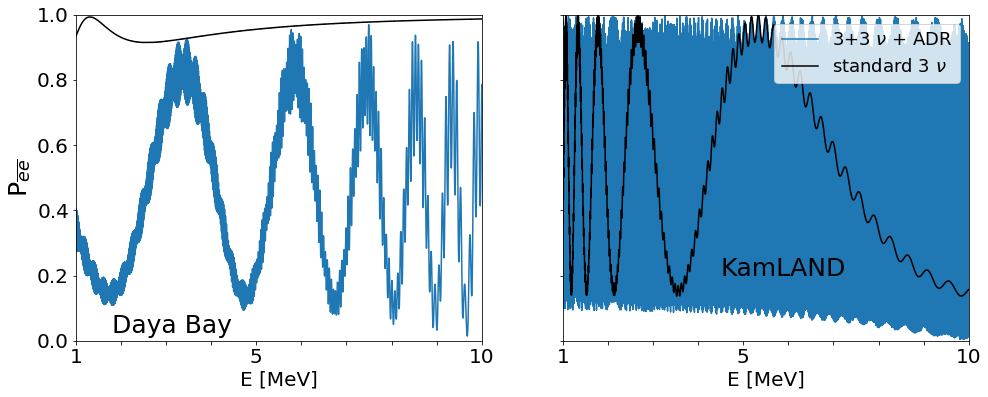}    
    \caption{Comparison between the predictions of the three neutrino standard picture and the 3+3 model with altered dispersion relations for Daya Bay (left) and KamLAND (right).}
    \label{fig:reac}
\end{figure}

Another problem arises in the calculation of the effective mass splittings. After diagonalizing the Hamiltonian, one can calculate the effective masses, $m^2_{i, \text{eff}}(E) = 2 E \lambda_i(E)$, and their differences from the eigenvalues, $\lambda_i(E)$, which depend on the energy. The desired behavior regarding the mass splittings is the following:
\begin{enumerate}
\item For energies below the one for which the LSND anomaly happens, $E < E_{\text{LSND}} \sim 10$ MeV, the standard mass differences $\Delta m^2_{21} $ and $\Delta m^2_{31}$ must be recovered. Therefore, the three
sterile neutrinos must be considerable heavier than the active ones.
\item For energies larger than the one for which the MiniBooNE excess is found, $E > E_{\text{MB}} \sim 300 $ MeV, the active and sterile neutrinos should decouple. In this range, $\Delta m^2 _{64}$ and $\Delta m^2 _{54}$ have to recover the values of $\Delta m^2 _{31}$ and $\Delta m^2 _{21}$, respectively.  
In these two regimes far away from the resonances, the active-sterile mixing angles must be small.
\item To explain the observed signals in  LSND and MiniBooNE, mass splittings of $\sim 1$ eV$^2$ are needed and the mixing angles have to be large.
\end{enumerate}
\begin{figure}[t!]
    \centering
    \includegraphics[width = 0.8\textwidth]{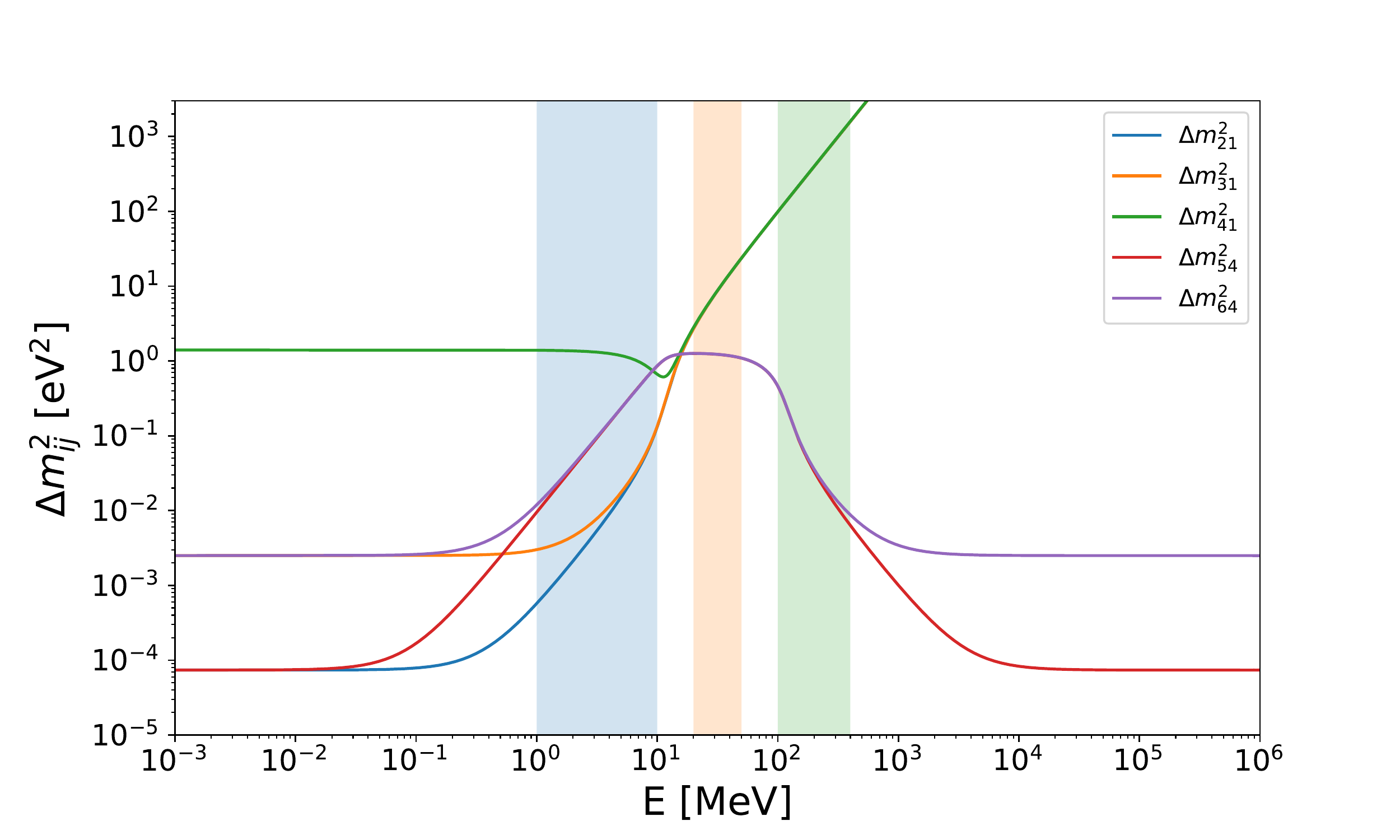}
    \caption{Effective mass splittings $\Delta m^2_{ij}$ as a function of the energy. The new parameters are fixed to the values from "set 2'' in Tab.~\ref{tab:newparam}. For the other set of parameters the picture looks very similar. The shaded region indicates the energy range relevant for reactor neutrino experiments (blue), LSND (orange) and MiniBooNE (green).}
    \label{fig:running}
\end{figure}
The energy dependence of the mass 
differences in this particular model is 
presented in Fig.~\ref{fig:running}. The 
resonant behavior needed to generate a 
large mass splitting for the energy ranges 
in LSND and MiniBooNE would also effect the 
energy range covered by reactor experiments 
(as indicated by the shaded regions in the 
figure), in particular Daya Bay and 
KamLAND, which observe neutrinos with 
energies in the range of $1-10$ MeV. Daya 
Bay can set strong constrains on this 
family of models, since its measurements of 
both $\theta_{13}$ and $\Delta m ^2_{31}$ 
are very accurate~\cite{Adey:2018zwh}. 
KamLAND, on the other hand, measured 
$\Delta m ^2_{21}$  with an excellent 
accuracy~\cite{Abe:2008aa}. In this energy 
range, the dependence of the mass 
splittings on the energy is very relevant. 
As it is shown in Fig.~\ref{fig:running}, 
the values of $\Delta m^2_{21}$ and 
$\Delta m^2_{31}$ predicted by the model 
differ significantly from the values 
measured at reactor experiments, namely 
$\Delta m^2_{21} \simeq 7.6\times 10^{-5}$ eV$^2$ and $\Delta m^2_{31} \simeq 2.5\times 10^{-3}$ eV$^2$. 
As a result, the predicted oscillation 
probabilities for Daya Bay and KamLAND  
deviate dramatically from the standard 
three neutrino framework, as one can see in 
Fig.~\ref{fig:reac}. Such a relevant 
deviation from the standard picture would 
have been easily detected already many 
years ago, so one can conclude that the 
model under study is not 
compatible with neutrino oscillation data.
%
%\sout{The choice of different values for the 
%parameters in the model does not change 
%this result qualitatively, so one can 
%conclude that this proposal for reconciling 
%LSND and MiniBooNE with other neutrino 
%oscillation data suffers from 
%incompatibilities with reactor neutrino 
%experiments. In the following subsection we 
%will focus on the case where only one 
%resonant effect is generated to explain the 
%MiniBooNE signal.} 
%
In the following  subsection, however, we will discuss a very particular case 
where all the three new parameters are rather small. This choice moves the resonant 
behavior away from the energy range  relevant to reactor neutrino experiments and, therefore, at low energies 
one recovers the effective 3+1 mixing. Hence, we can satisfactorily
explain KamLAND and Daya Bay, since they are mostly unaffected by the 3+1 mixing, as well as very short baseline reactor experiments and LSND.
However, as we will see, the tensions with T2K will be still present.

%%%%%%%%%%%%%%%%%%%%%%%%%%%%%%%%%%%%%%%%%%%%%%%%%%
%%%%%%%%%%%%%%%%%%%%%%%%%%%%%%%%%%%%%%%%%%%%%%%%%%
%%%%%%%%%%%%%%%%%%%%%%%%%%%%%%%%%%%%%%%%%%%%%%%%%%
\subsection{A resonant explanation for MiniBooNE avoiding reactor constraints}

As we have seen, the explanation of the MiniBooNE and LSND anomalous signals using resonant effects is in very strong tension with the well established reactor experiments. However, there is a conceptually interesting possibility that arises as a  modification of the initial proposal~\cite{Doring:2018cob}.
If the three parameters in Eq.~\ref{eq:V3+3} ($\epsilon$, $\kappa$ and $\xi$) have similar values in a range such that the corresponding resonant energies lie in the region where the excess of events is found in MiniBooNE, one can avoid the inconsistencies with reactor experiments. This is due to the fact that, if the resonant effect happens at the order of $\mathcal{O}(0.1 \text{GeV})$, the energy dependence of the mass splittings will not manifest in the energy range of the reactor experiments. The overall behavior in this particular case would be the following:
\begin{itemize}
\item For energies below $\sim$ 100~MeV, neutrino oscillations would be described by an effective 3+1 picture. This allows to accommodate the LSND signal while being consistent with reactor experiments. 
\item At  $\sim$100-500~MeV, a resonant effect would account for the anomalous signal found in MiniBooNE.
\item At higher energies, as it was discussed before, one would recover the three-neutrino picture once the parameters are chosen ad hoc to reproduce the experimental results. In this case, bounds from long baseline experiments would not apply directly to the parameters of the effective 3+1 picture at low energies.
\end{itemize}

\begin{figure}[t!]
    \centering
    \includegraphics[width = 0.8\textwidth]{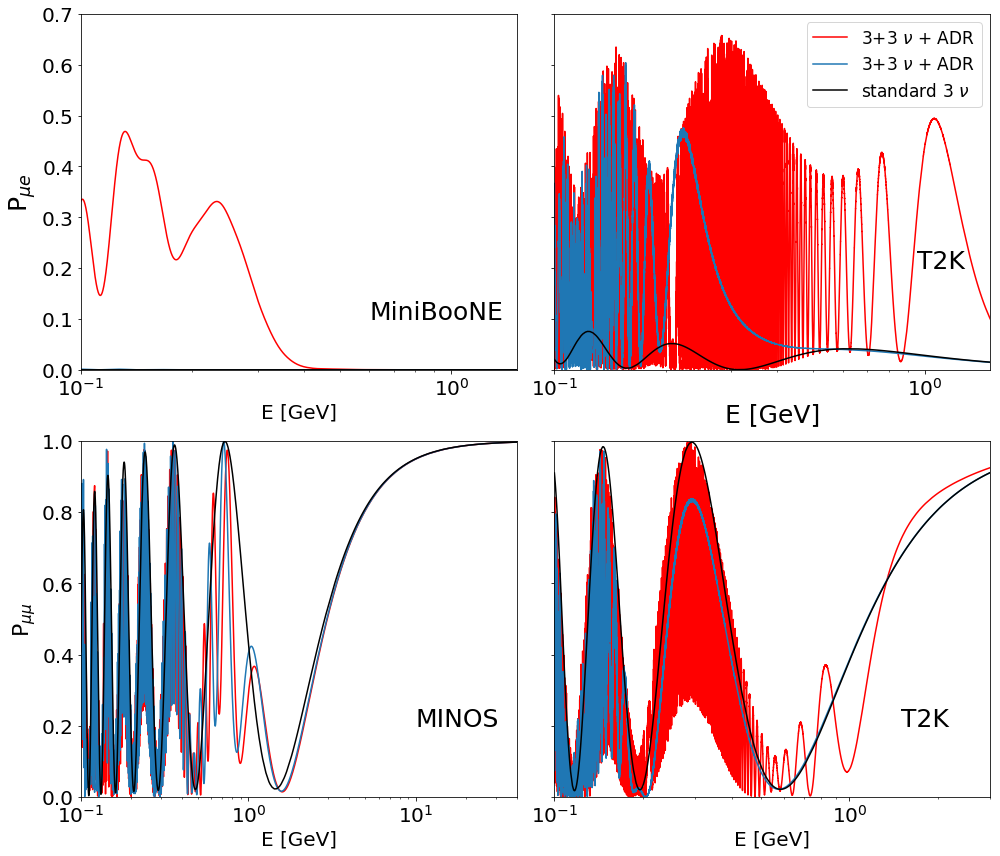}
    \caption{Comparison between the predictions of the three-neutrino standard picture (black) and a 3+3 model with altered dispersion relations for  $(\epsilon, \kappa, \eta) = (4.9, 5, 5)\times 10 ^{-17}$ (blue) and $(\epsilon, \kappa, \eta) = (1, 5, 5)\times 10 ^{-17}$ (red). } 
    \label{fig:one_res}
\end{figure}

Nonetheless, predictions for experiments in the energy range between 100~MeV and 10~GeV are expected to be modified after considering these altered dispersion relations.  The impact is expected to be particularly large in T2K, as previously shown. Deviations would appear in MINOS too, when the values of the parameters are chosen to explain the MiniBooNE signal with this mechanism, as is shown in Fig.~\ref{fig:one_res}. 
There, we show the predicted probabilities at MiniBooNE, MINOS and T2K for two different choices of parameters, as indicated in the caption.
As can be seen in the figure, small deviations 
from this fine tuned scenario  can also wreck 
the desired behavior of the oscillation 
probability in MiniBooNE, as indicated by the 
blue line. Note that this line in the 
MiniBooNE panel is systematically very close 
to zero and therefore this scenario would 
not create an excess. We show this scenario to 
highlight the instability of these solutions 
meant to avoid the reactor constraints and to 
show the level of fine-tuning needed to find 
them.

In addition to the 
discussion presented at the probability level, we have also  calculated a $\chi^2$ value for our red
benchmark point. Note that the blue  benchmark point is already excluded by 
MiniBooNE, since it does not produce a  sizeable oscillation probability there. 
This test has been performed using the same T2K  data~\cite{Abe:2018wpn} and the same statistical
analysis as in  Ref.~\cite{deSalas:2017kay}. To calculate the $\chi^2$ value, we marginalize over 
all the standard oscillation parameters  relevant for T2K, namely $\Delta m_{31}^2$, $\sin^2\theta_{23}$, $\sin^2\theta_{13}$ and $\delta$. We 
obtain a value of $\chi^2\approx 237$, to  be compared to the value in the standard  neutrino oscillation scenario $\chi^2 \approx 120$, for 102 degrees of freedom. 
We should also remark that, in the scenario with ADRs, the  best fit value for $\sin^2\theta_{13}$ 
turns out to be very small. Then, if we include in our analysis 
a prior on $\sin^2\theta_{13}$ coming from the Daya Bay measurement,  the minimum $\chi^2$ value increases further to 
$\chi^2\approx 285$. Note that imposing a  prior in this case is very well justified, since 
the new oscillation parameters were  chosen to have no effect on reactor 
neutrino experiments. Thus, although our benchmark point  would give rise to a significantly large 
probability that could potentially explain MiniBooNE, it is ruled out by 
T2K. A systematic search for points simultaneously compatible with the observed signals in MiniBooNE and T2K has produced no result. Indeed,
a similar behavior to the one described above can be observed for any other point producing an observable oscillation probability at MiniBooNE: they are penalized 
with huge $\chi^2$ values in T2K.
Therefore, one can conclude that this hypothesis does not provide a satisfactory explanation of neutrino oscillation data, including the anomalous LSND and MiniBooNE signals.

%%%%%%%%%%%%%%%%%%%%%%%%%%%%%%%%%%%%%%%%%%%%%%%%%%%%%%%%%%%%%%%%%%%%%%%%%%%%%%%%%%%%%%%%%%%%%%%%%%%%%%%%%%%%%%%%%%%%%%%%%%%%%%%%%%%%%%%%%%%%%%%%%%%%%%%%%%%%%%%%%%%%%%%%%%%%%%%%%%%%%%%%%%%%%

\section{Conclusions}
\label{sec:conc}
We have shown that an additional sterile neutrino satisfying an altered dispersion relation arising as a consequence of an effective potential can not give an explanation of the MiniBooNE signal while, at the same time, being consistent with long baseline experiments, mainly MINOS/MINOS+ and T2K.
Even in the case of more complex models with additional sterile neutrinos, the modification of the dispersion relation  can not explain  in a consistent picture current neutrino oscillation data and the observed anomalies. 
First, one finds that the resonant mixing angles required to explain the LSND and MiniBooNE excesses  would have given rise to signals in other experiments, unobserved so far. Moreover, the dependence of the effective mass squared differences on the energy is strongly constrained by current reactor data and in disagreement with the predictions of this type of models. It is actually possible to avoid the constraints from reactor experiments if the resonant behavior is only invoked to explain the MiniBooNE anomalous signal. In this case, the lowest energy observables (essentially reactor  and LSND data) will be described by an effective 3+1 scenario, free of any further constraints from higher-energy experiments. Nevertheless, this proposal requires high levels of fine tuning and is very disfavoured by T2K results.
Therefore, sterile neutrinos with altered dispersion relations can be added to a growing list of better or worse motivated physics that can not explain the anomalies observed in neutrino oscillation experiments.
Should one come up with a model including any form of altered dispersion relations, these two energy-dependent effects have to be correctly addressed, since they would set strong constraints in the parameters of the model under study. 

Models of great complexity can be built in order to seek for an explanation to the anomalies in terms of sterile neutrinos. However, the number of parameters they require grows rapidly. The spirit that led to the proposal of oscillations with sterile neutrinos was to keep the explanation simple. If a large number of parameters was needed to phenomenologically explain the results from all the experiments, there would be no point on talking about sterile neutrino oscillations, since one would be eventually parametrizing some other physical phenomena. Therefore, greater efforts should be made in the search for explanations of the LSND and MiniBooNE signals which are not related to oscillations into sterile neutrinos.

As a parting remark, we would also like to mention that models with extra neutrinos can be seriously challenged by cosmological limits on the additional number of relativistic degrees of freedom, depending on the specifics of the dispersion relation. Likewise, a resonant mixing at the MeV scale can be severely compromised by BBN results.

\section*{Acknowledgments}
We would like to thank Jorge D\'{i}az and Heinrich P\"{a}s for useful comments and discussions.
GB acknowledges support from  the MEC and FEDER (EC) Grant SEV-2014-0398, FIS2015-72245-EXP, and FPA2017-845438 and the Generalitat Valenciana under grant PRO\-ME\-TEOII/2017/033, also partial support from the European Union FP10 ITN ELUSIVES (H2020-MSCAITN-2015-674896) and INVISIBLES-PLUS (H2020- MSCARISE- 2015-690575).
PMM, CAT and MT are supported by the Spanish grants FPA2017-90566-REDC (Red Consolider MultiDark), FPA2017-85216-P and
SEV-2014-0398 (MINECO/AEI/FEDER, UE), as well as PROMETEO/2018/165 (Gene-ralitat Valenciana). 
PMM acknowledges financial support from MICINN through Programa Estatal de Fomento de la Investigación Cient\'{i}fica y T\'{e}cnica de Excelencia - Centros Severo Ochoa y Unidades Mar\'{i}a Maetzu - CSIC and the FPU grant FPU18/04571.
CAT is supported by the FPI grant BES-2015-073593. PMM and CAT receive partial support from the EU Horizon 2020 project InvisiblesPlus (690575-InvisiblesPlus-H2020-MSCA-RISE-2015). PMM and CAT are grateful for the kind hospitality received at Fermilab during the final stage of this work.
MT acknowledges financial support from MINECO through the Ram\'{o}n y Cajal contract RYC-2013-12438.

\section*{References}
%\bibliography{bibliography}

\end{document}